\documentclass[traditabstract]{aa}
\usepackage{graphicx}
\begin{document}

\title{Stellar metallicity of star-forming galaxies at $z\sim3$
	\thanks{Based on ESO observations, proposals 082.A-0398 and 084.A-0367 }}

   \subtitle{}

   \author{Veronica Sommariva
          \inst{1}
          \and Filippo Mannucci \inst{1,2}
          \and Giovanni Cresci \inst{1}
          \and Roberto Maiolino \inst{3}
           \and Alessandro Marconi\inst{4}
          \and Tohru Nagao \inst{5}
          \and Andrea Baroni  \inst{4}
          \and Andrea Grazian \inst{3}
          }

   \institute{INAF-Osservatorio Astrofisico di Arcetri, Firenze, Italy  \\
     \thanks{}
              \email{veronica@arcetri.astro.it}
              \and
            Harvard-Smithsonian Center for Astrophysics, 60 Garden street, 
            Cambridge, MA 02138, USA
            \and
         INAF-Osservatorio Astronomico di Roma, Monte Porzio Catone, Italy
           \and
         Dipartimento di Astronomia, Universit\`a di Firenze, Firenze Italy
         \and
         Kyoto University, Japan
             \\
             }

   \date{Received 22-09-11; accepted 09-12-11}


  \abstract{
  The stellar metallicity is a direct measure of the amount of metals present in a galaxy,
 as a large part of the metals lie in its stars.
In this paper we investigate new stellar metallicity indicators suitable for high-z galaxies studying 
the stellar photospheric absorption lines in the rest frame ultraviolet, hence sampling predominantly 
young hot stars.
We defined these new indicators based on the equivalent widths (EW) of selected features using theoretical 
spectra created with the evolutionary population synthesis code {\it Starburts99}. 
We used them to compute the stellar metallicity for a sample  of UV-selected galaxies at $z>3$ from the AMAZE survey
 using very deep (37h per object) VLT/FORS spectra. Moreover, we applied the new metallicity indicators
to  eight additional high redshift galaxies found in literature. We then compared 
stellar and gas-phase metallicities measured from the emission lines 
for all these galaxies, finding that within the errors 
the two estimates are in good agreement,  
with possible tendency to have stellar metallicities lower than the gas phase ones. 
For the first time, we are able to study the stellar mass-metallicity relation at $z>3$. 
We find that the metallicity of young, hot stars in galaxies at $z\sim3$ have 
similar values of the aged stars in local SDSS galaxies, 
contrary to what observed  for the gas phase metallicity.
}


 \maketitle
%

\section{Introduction}

Metallicity is one of the important properties of galaxies, and
its study is able to shed light on the details of galaxy evolution.
It is, in fact, an integrated property, 
related to the whole past history of the
galaxies.
In particular, metallicity is sensitive to 
whole star formation history, and so to 
evolutionary stage of the galaxy. 
Moreover, it is affected by presence of infalls and outflows, i.e.
by feedback processes and by the interplay between the forming galaxy 
and the intergalactic medium (see e.g. Erb et al. 2008, Mannucci et al. 2009, Cresci et al. 2010).
As consequence, it has become an important test of galaxy evolution (e.g.
Nagamine et al. 2001, Spitoni et al. 2010, Dav\`e et al. 2011).\\
Local galaxies show a clear correlation between mass and metallicity (MZR),
for which the galaxies with larger 
stellar mass have higher metallicities,
and this correlation appears to hold both in term of gas-phase metallicity 
(e.g. Tremonti et al. 2004) and stellar metallicity (Gallazzi et al. 2005, Panter et al. 2008).\\

At high-redshift the gas-phase metallicity of the ISM of star-forming galaxies
has been measured using primarily oxygen abundances. 
The most common techniques to determine the gas phase metallicity 
are based either on theoretical calibrations (see Kewley \& Ellison 2002, and 
Kewley \& Ellison 2008) or
on empirical metallicity calibrations, the so-called
``strong line diagnostics'', which are based on the ratios of collisionally
excited forbidden lines to hydrogen recombination lines.
Previous studies have shown that the mass-gas phase metallicity relation
presents evidence of strong redshift evolution.
Among others, Savaglio et al. (2005) and Zahid et al. (2011) studied  star forming galaxies
at redshift  $z\sim0.7$ and demonstrated that, at given 
mass, these galaxies shown lower metallicity than the SDSS sample
at $z\sim0.1$. Erb et al. (2006) reported a more significant
decrease of metallicity in galaxies at $z\sim2.2$.
Two projects were specifically designed to extend the investigation
of MZR at  $z>3$: LSD (Lyman-break Stellar 
population and Dynamic) and AMAZE (Assessing the Mass-Abundance 
redshift Evolution).
With these projects, Maiolino et al. (2008) and Mannucci et al. (2009)
showed  for the first time the evolution of the mass-metallicity
relation at $z>3$.
However, the redshift evolution of the gas phase  metallicity in galaxies 
have been questioned recently by Mannucci et al. (2010).
They discovered that metallicity depends not only on the mass,
but also from the Star Formation Rate (SFR): for a given stellar mass, 
galaxies with higher SFRs systematically show lower metallicities.
This is the so-called ``Fundamental Metallicity Relation (FMR)'', i.e.,
a tight relation between stellar mass, gas-phase metallicity, and
star formation rate (SFR). Local SDSS galaxies 
show very small residuals around this relation, of the order of 0.05dex.
Yates et al. (2011) found a similar
relation, with some differences due to the metallicity calibration adopted.
According to Mannucci et al. (2010), the FMR does not appear to evolve
 with redshift up to z$\sim$2.5, with the high redshift
galaxies following the same FMR defined by the local SDSS galaxies.
This suggests that the observed evolution of the mass-metallicity relation
is due to selection effects and to the increase of the average SFR with redshift.
In fact, the measured metallicity in several additional samples of high-z galaxies 
results to be in agreement with the predictions of the FMR given the mass and SFR of the galaxies:
galaxies having lower SFRs than the general population at their redshifts also have higher metallcities (e.g., Richard et al. 2011, Nakajina et al. 2011), 
and galaxies with higher SFRs also have lower metallicities (e.g., Erb et al. 2010, Contini et al. 2011, Sanders et al. 2011), 
so that all these galaxies follow the FMR.
Also, the FMR allowed Mannucci et al. 2011 and Campisi et al. 2011 to show that the hosts of the long-
GRBs have the same metallicity properties of the other star-forming galaxies.
However, they found some metallicity evolution of the FMR at $z\sim3.3$, where
galaxies tend to have lower metallicities.\\

All the observational studies mentioned in the previous  paragraph refer to the gas-phase metallicity,
as measured by emission lines.
Gallazzi et al. (2005) presented the local mass-stellar metallicity relation based
on $\sim170000$ SDSS galaxies (Sloan Digital Sky Survey Data Release Two). 
The stellar metallicities were derived using the Lick system of spectral indices in the 
optical region which are sensitive to the overall metallicity of the stellar population,
primarily dominated by intermediate/old stars.
They adopted a Bayesian statistical approach and  derived the stellar metallicities
by comparing the observed spectrum of each galaxy with a comprehensive library of model 
spectra corresponding to different star formation histories. 
They found that at  low  masses, the stellar metallicity increase with  mass,
while above $\sim3\times10M_{\odot}$ the relation flattens out. 
In addiction they noted that gas-phase metallicity is best determined for star-forming galaxies,
 whereas stellar metallicity is best determined for early-type galaxies, and found that
the stellar metallicity is generally lower than the gas-phase metallicity (by ∼0.5 dex).
More recently Panter et al. (2008) inferred the stellar metallicity history of SDSS galaxies
and determined their stellar mass-metallicity relation. 
They used a different approach respect to Gallazzi et al. (2005), but they found similar results. 
Moreover,  considering only the younger  population of galaxies ($\leq$1 Gyr) they found good agreement 
also with the gas phase metallicities. \\

A very  limited work has instead been done on stellar metallicity at high
redshift, see Shapley (2011) for a recent review.
As in local {\it starbursts}, the strongest features in the rest-frame
UV spectrum of distant galaxies are interstellar and photospheric absorption 
lines of C, N, O, Si, and Fe, produced by  hot, young O-B stars  (see Shapley et al. 2003).
One advantage in using these spectra to measure the metallicity at $z\sim3$ is that
the UV rest frame is shifted into the optical spectral region, which is
easier to observe from the ground-based telescope. 
However, very high signal to noise on the stellar continuum is required 
to study the relevant absorption features for metallicity measurements,
 and therefore such studies were obtained, until now, mainly
for gravitationally lensed galaxies(Rix et al. 2004, Quider et al. 2009, 
Dessauges-Zavadsky et al. 2010) or for co-added star forming galaxy spectra (Halliday et al. 2008).\\

Several authors have presented calibrations for stellar metallicity
based on UV absorption lines.
  Leitherer et al. (2001)
investigated the influence of metallicity on the spectra of 
star forming galaxies. 
In particular, they investigated the existence of some 
blended photospheric lines whose strengths depend on metallicities only.
They found that the two blends of lines near $\lambda$1370
and $\lambda$1425 
(which they attributed to OV $\lambda$1371 and FeV $\lambda$1360-$\lambda$1380
 and to $\lambda$SiIII 1417, CIII $\lambda$1427  and FeV $\lambda$1430
respectively) have equivalent widths that increase steadily 
with metallicity and do not depend on other stellar parameters, as age and IMF.
Rix et al. (2004) using the {\it Starburst99} plus their 
non-LTE model atmosphere code WM-basic, 
supported the conclusions
that these lines are useful metallicity indicators, 
and suggested a 
new indicator at $\lambda$ 1978.
Rix et al. (2004) applied the new indicators 
to measure the stellar metallicity  of two lensed galaxies,
MS 1512-cB58 at z=2.73, and Q1307-BM1163 at z=1.411, finding good agreement with the gas phase metallicity.\\

On the basis of these works, Mehelert et al. (2006) calculated the EWs of the 
 $\lambda$1370  $\lambda$1425  for 12 galaxies
with $2.37<z<3.40$ in the FORS deep field. They investigated the evolution of the
EWs and the metallicity with the redshift, and found that the abundances of heavy elements
is increasing between $z\sim0.5$ and   $z\sim2.5$.
Halliday et al. (2008), instead,  computed the stellar metallicity using the  $\lambda$1978
 indicator 
 for a co-added unlensed star forming galaxies spectra at $z\sim2$.
Comparing their results with the gas phase metallicity of galaxies with similar
mass, they found that their stellar metallicities were lower by a factor of $\sim0.25dex$.  
Quider et al. (2009) computed the stellar metallicity for the gravitationally lensed
galaxies Cosmic Horseshoe at $z=2.38$ by using the $\lambda$1425  metallicity indicator,
found good agreement between stellar and gas phase metallicity.
More recently, Quider et al. (2010) carried on the same work for the lensed galaxies
Cosmic Eye at $z=3.07$. Because of the presence a strong sky lines in the region of the stellar metallicity
indicators, they do not give the exact value of the stellar metallicity, but just an indication
by comparing the observed spectra with the model.
Finally, Dessauges-Zavadsky et al. (2010) studied the stellar metallicty for the lensed galaxy 8 o'clock arc
at  $z=2.73$. Using the  $\lambda$1425 and  $\lambda$1978 photospheric indices, 
they found comparable metallicity for the gas and the stellar component.\\

All these studies provide important constraints 
to the models of galaxy formation.
In fact, many models study  the processes of gas infall and outflows, 
and  the properties of feedback and galactic winds
(Dekel \& Woo 2003, 
Veilleux, Cecil \& Bland-Hawthorn 2005, 
Murray, Quataert \& Thompson 2005, 
Dav\'e, Finlator \& Oppenheimer 2007,
de Rossi, Tissera \& Scannapieco 2007, 
Brooks et al. 2007, 
Tornatore et al. 2007,
Somerville et al. 2008,
Oppenheimer \& Dav\'e 2008, 
Finlator \& Dav\'e 2008, 
Mouchine et al. 2008,
Dayal et al. 2009,
Tassis, Kravtsov \& Gnedin 2008, 
Kobayashi et al 2007, Calura et al. 2009,
Spitoni et al. 2010, 
Salvaterra et al. 2010,
Sakstein et al. 2011).
The no evolution of the FMR, the associated evolution of the 
mass-metallicity relation, and the properties of the effective yields 
put strong constraint on them. 
In this picture, the stellar metallicity can be used 
as independent measurement to test the predicted metallicity evolution
(De Rossi et al. 2007, Dav\'e \& Oppenheimer 2007).\\

In this work we study the stellar metallicity of a sample of  galaxies at $z>3$.

In the next section we review the different indices used for stellar metallicity estimates at high-z, 
and define new rest frame UV feature suitable for such measurements  to enlarge the number of available
metallicity indicators. 
In Section 3 we present our sample of high redshift galaxies and
we compute their stellar metallicity applying the calibrations found.
In Section 4 we calculate the stellar and gas phase metallicity for some 
lensed galaxies found in literature, and we compare the two. 
Finally we present the first stellar 
mass-metallicity relation obtained at $z>3$, followed by the conclusions.


\begin{figure*}
   \centering
   \includegraphics[width=12cm,angle=270]{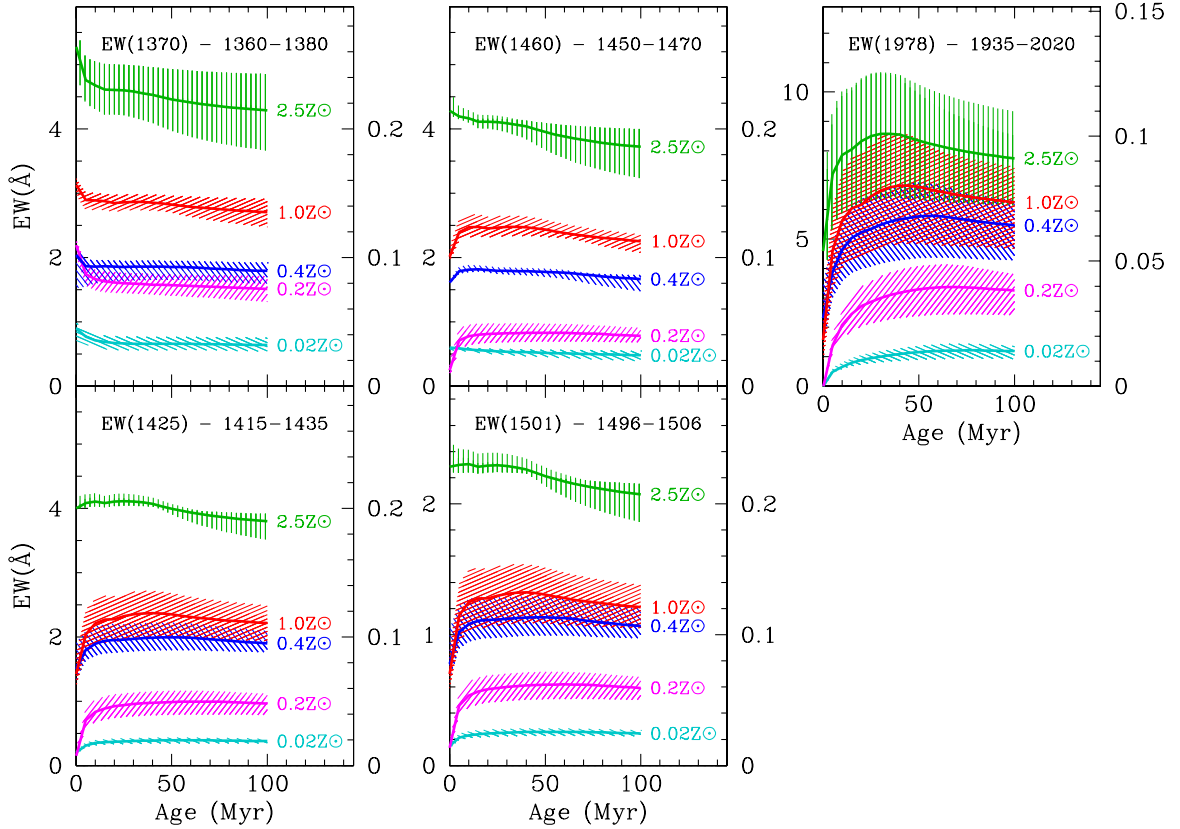}
      \caption{
        Variation of the line indices with stellar population age and  metallicity.
        In each plot, the EW is shown in left vertical axis, while the right
        axis shows the average fractional depth below the continuum.
        For each feature we draw the models constructed with   {\it Starburts99}
        at five value of metallicity as shown in different colors.
        For each metallicity, the color regions represent the error due to the dependence on the 
        IMF assumed.
        In this way we can highlight the dependence of the models and robustness of the indices
        from the IMF and stellar population age assumed.
        }
         \label{linvar}
   \end{figure*}
 
\section{Stellar metallicity from absorption lines}
\label{sec:calibintro}

When studying high-redshift galaxies, is useful to have a large 
number of calibrated features over a large wavelength range to increase 
the constraints on metallicity and avoid 
the effects of atmospheric absorption 
bands and bright sky emission lines.

A feature can be used as a metallicity indicator if the following 
conditions are satisfied: 
1) the EW is deep enough to be measured in high redshift galaxies;
2) it varies significantly with the metallicity; and
3) it does not depends critically on age or IMF.

The aim of this work is to update the old metallicity calibrations
and to enlarge the number of indices over all the range of UV spectral region.\\
We want measure the stellar metallicities of 
high redshift galaxies by comparing the observed
photospheric lines to the results of the
population synthesis code  {\it Starburts99} (Leitherer et al. 1999). 
The original version of this code allowed the creation
of synthetic UV spectra with a variety of ages and IMFs,
and used the stellar evolution models to follow
the stellar population over the time.
The first release included only few different metallicities because empirical
stellar libraries were available only for Milky Way
O-type stars observed with the International Ultraviolet
Explorer (IUE) satellite.
The galactic B stars were later included to that library 
by de Mello et al. (2000). In order to consider the sub-solar
heavy elements abundences, a new improvement was obtained in 2001
 with the inclusion of a library of O-type
spectra obtained from HST STIS observations of stars
in Large and Small Magellanic Clouds  (Leitherer et al. 2001).
Rix et al. (2004) followed an other approach  by utilizing 
theoretical library spectra instead than empirical ones.
Their purpose was synthesized the photospheric 
absorption lines seen in the spectra of star forming galaxies
taking into account on the 
non-LTE model atmosphere, the  stellar wind
and their effects in the spectral synthesis of hot stars.
Therefore they replaced the empirical library with a grid
of theoretical spectra generated with the hydrodynamics code  WM-basic. 
The major difference between this two approaches
is that the empirical spectra include prominent UV interstellar
absorption lines not present in the theoretical 
spectra, which are purely stellar. Nevertheless, they found good 
agreement between the empirical and the theoretical spectra.

Because Rix et al. (2004) in principle
focus on photospheric lines, processes such as shock
emission that affect only the high-ionization wind lines were not
included in their models. These limitations have been addressed
to some extent in the latest generation of the WM-basic code
(Leitherer et al. 2010),  which is  used in this work.
The latest generation
of the WM-basic code is optimized to compute the strong P Cygni type
lines originating in the wind of the hot stars. 
This is a great advantage  respect use only 
the faint photospheric lines, because the stellar-wind features are stronger, 
therefore more easy to detect 
in low S/N spectrum.

{\it Starburst99} with the WM-basic library allows the creation of simulated spectra 
depending on a number of free parameters related to star 
formation history, IMF, age, metallicity, supernova and black
hole cut-off, stellar atmospheres and microturbolence. 
We generated galaxy model spectra using the Padova tracks,
with thermally pulsing AGB stars added, for five values 
of metallicities, 0.02, 0.4, 0.2, 1.0 and 2.5 $Z_{\odot}$,
and assuming continuous star formation histories, results in Fig.~\ref{linvar}
We considered five different stellar initial mass function (IMF).
The reference model is a  classical Salpeter power law with exponent
$\alpha=2.35$ and upper mass limit $M_{up}=100M_{\odot}$.
We also  considered IMFs with
$\alpha=1.85$  and $\alpha=2.85$  between $1M_{\odot}$ and
$100M_{\odot}$, and  with $\alpha=2.35$ and  mass limit $M_{up}=60M_{\odot}$.
Finally, we compute models using the Kroupa IMF.

\begin{figure*}
   \centering
   \includegraphics[width=13cm, angle=270]{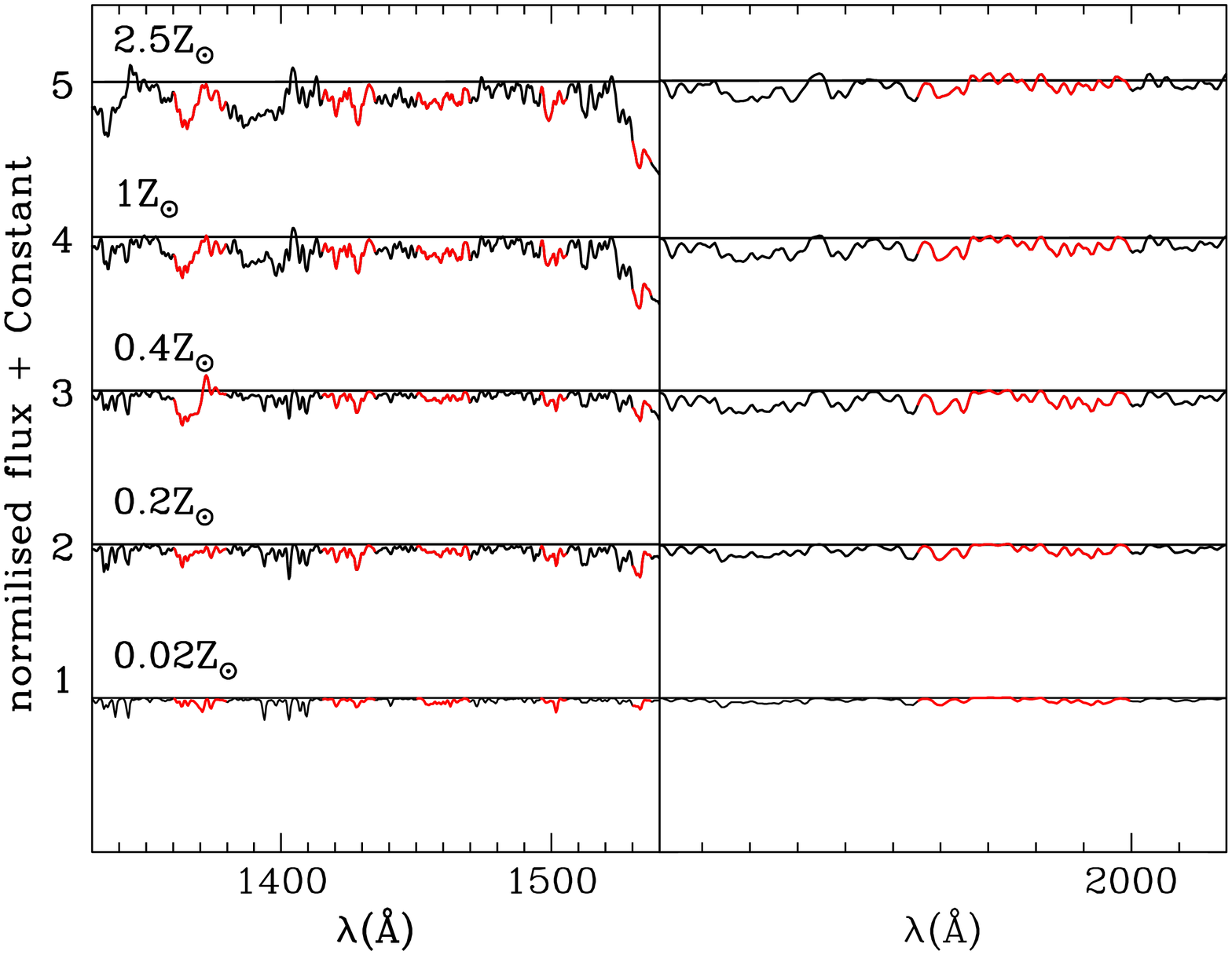}
      \caption{Smoothed and normalized reference models
obtained with   {\it Starburts99}  at metallicity 0.02, 0.4, 0.2, 1 and 2.5$Z_{\odot}$.
In red are highlighted the metallicity indicators discussed in the text to show
their dependence  from metallicity.
              }
         \label{mod}
   \end{figure*}

\begin{table}
\caption{List of the metallicity indicators with the corresponding
elements, and regions where the equivalent
widths are integrated. For an additional indicator at $\lambda$1533 see sec.~\ref{1533}}   
\label{IND}      
\centering      
\begin{tabular}{ccc}     
\hline\hline       
Indicator ID & element &  $\lambda$range \\ 
\hline                    
F1370  & OV, FeV          & 1360$-$1380\tablefootmark{a}\\  
F1425  & CIII, FeV, SiIII & 1415$-$1435\tablefootmark{a}\\
F1460  & NiII             & 1450$-$1470\tablefootmark{b}\\
F1501  & SV               & 1496$-$1506\tablefootmark{b}\\
F1978  & FeIII            & 1935$-$2020\tablefootmark{c}\\
\hline                  
\end{tabular}
\tablefoot{  
\tablefoottext{a}{Leitherer et al. 2001 }
\tablefoottext{b}{this work} 
\tablefoottext{c}{Rix et al.2004}.
}
\label{list} 
\end{table}


\subsection{Metallicity indicators}
\label{indic}

The UV spectrum of star forming galaxies is dominated by strong 
photospheric absorption features that are sensitive to metallicity.

Leitherer et al. (2001) investigated the existence of some 
blended photospheric lines whose strengths depend on metallicities only.
They found that the two blends of lines near $\lambda$1370
and $\lambda$1425 
(which they attributed to OV $\lambda$1371 and FeV $\lambda$1360-$\lambda$1380
 and to $\lambda$SiIII 1417, CIII $\lambda$1427  and FeV $\lambda$1430
respectively) have equivalent widths that increase steadily 
with metallicity and do not depend on other parameters.
Rix et al. (2004) supported the conclusions
that these lines are useful metallicity indicators, 
and suggested a 
new robust indicator at $\lambda$ $\sim1978$ \AA.

\begin{figure*}
   \centering
\includegraphics[clip=,width= .49\textwidth]{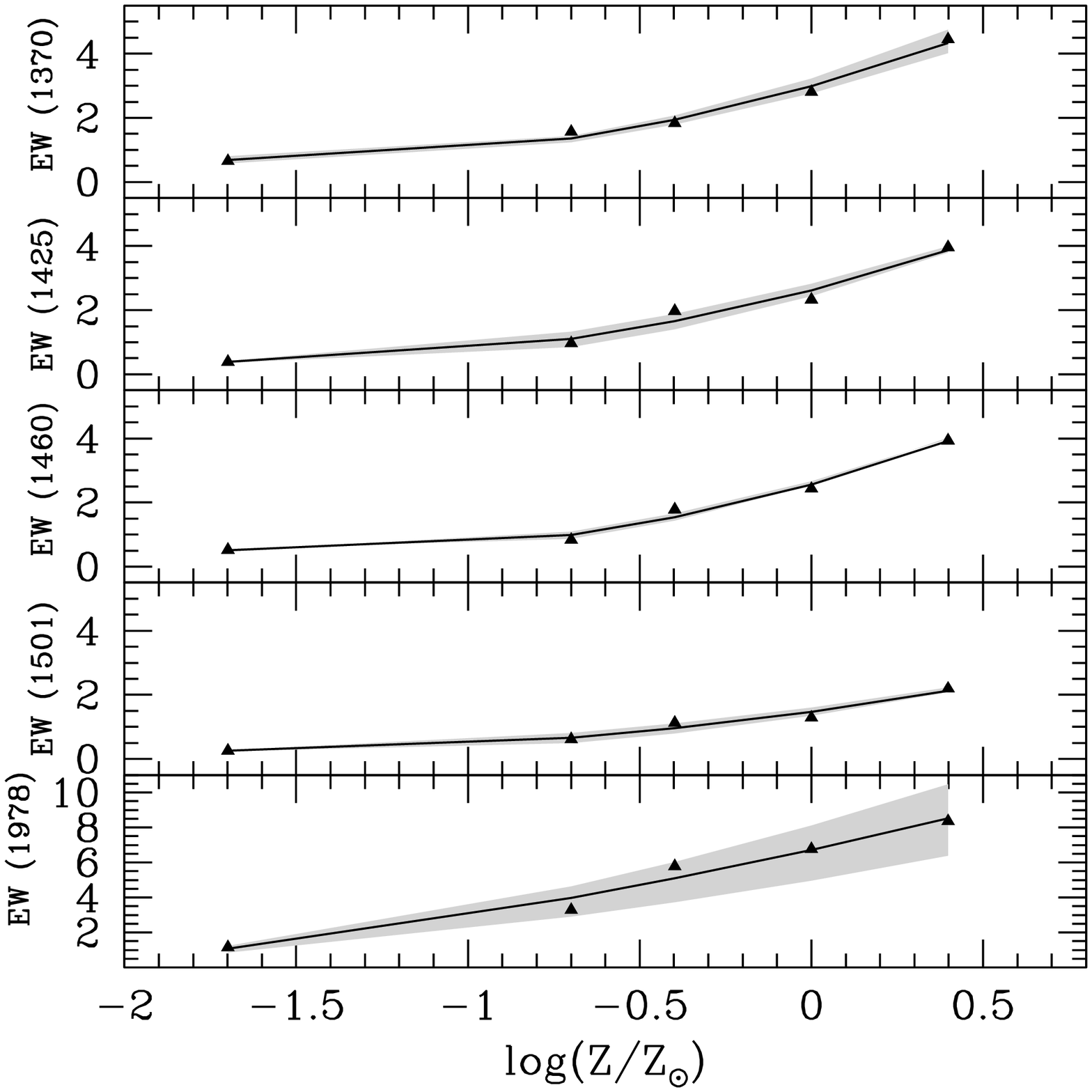}
\includegraphics[clip=,width= .49\textwidth]{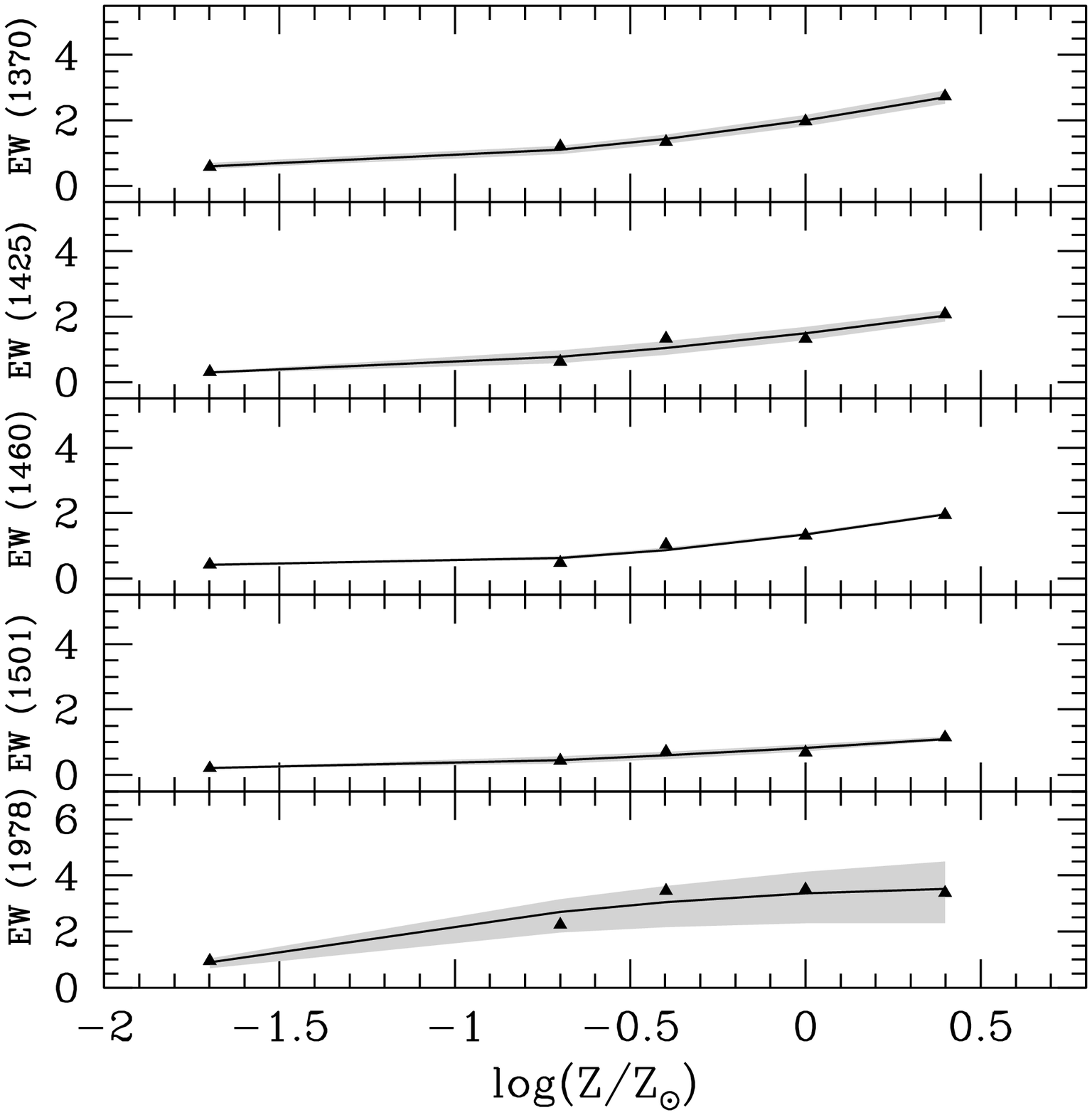}

      \caption{EW-Z relations for the five indices considered. The triangles represent equivalent widths measured 
        from the reference models at age 50 Myr, while the solid line is the
        quadratic fit given by equations in Table~\ref{tab:calibrations} for the real
        continuum (left panels) and Table~\ref{tab:pscal} for the ``pseudo-continuum''
        (right panels). The grey region
        represents the error due to the dependence of the models on the IMF assumed.
              }
         \label{cal1}
     \end{figure*}

Here we use the spectra of the new version of {\it Starburst99}
to update the calibrations of the features mentioned above and find 
additional new  useful features.
Using the new library, our measurements confirm 
that the indicators proposed in the  previous studies,
(F1370, F1425, and F1978), 
are stable after $\sim30Myr$ from the onset of star
formation, and increase monotonically with the metallicity 
with mild dependence on other parameters (see Fig.~\ref{linvar}).
Previous works have successfully used  these indices 
(Halliday et al. 2008 and  Quider et al. 2009).
The predicted EWs derived using the last version
of the  {\it Starburst99} code are similar
from the previously ones used for the 1425 index (0.11\AA~at solar metallicity).
However for the 1978 index the difference is higher, especially at high metallicity
(1.95\AA~at solar metallicity).
Moreover, the F1978 index is quite sensitive to the assumed IMF 
(see Fig. \ref{cal1}), making it more uncertain among the metallicity indicator.\\

\subsection{New indicators}
\label{ind}

The first region that we investigate is between 1496 and 1506 \AA~. 
We choose this region because the SV $\lambda$1501 line is an absorption
feature that arise in the photosphere of the hot stars, as 
Pettini et al. (1999)  and Quider et al. (2009) mentioned.
Fig.~\ref{linvar} shows  the age stability and the dependence
on metallicity of this index: as for the previous cases, this
line satisfies the necessary conditions to be used as indicator.

Another region that we consider is between 1450 and 1470 \AA~.
No one mentioned before this region for some photospheric lines,
but we found that the NiII at  $\lambda$1460 feature has the same behavior 
of the other lines discussed above, with the equivalent
width depending strongly on the metallicity, but not 
on age and IMF.\\
We therefore defined two new line indices, F1460, and F1501,   
as the equivalent widths integrated between 1450-1470 \AA~
and 1496-1506 \AA~ respectively.\\ 

The list of the metallicity indicators presented above 
are reported in  Tab.~\ref{list}.
Fig. ~\ref{mod} presents the smoothed and normalized reference models
obtained with   {\it Starburts99}  at metallicity 0.02, 0.4, 0.2, 1, and 2.5$Z_{\odot}$.
The strengths of the absorption features discussed above is clearly a 
strong and monotonic function of metallicity.

\subsection{Metallicities calibrations}
\label{mc}

For an accurate determination of  metallicity,
we interpolate the relation between the equivalent widths
of the indicators as a function of $Log(Z/Z_{\odot})$ discussed above.\\
We fitted a second-order polynomial to measured the values of the equivalent widths.
The results of the fitting is presented in Fig.~\ref{cal1}:
the  triangles represent the reference model (Salpeter IMF with 
$\alpha=2.35$ and age$=$50 Myr), the black line is the quadratic fit of 
these models, and the grey region represents the  uncertainty  on the
calibration due to the different IMF assumed, as shown in Fig.~\ref{linvar}.\\
We use a second-order fit expressed as:
 \begin{equation}
 \label{equ}
log(Z/Z_{\odot})= \alpha + \beta EW + \gamma EW^2
\end{equation}

The coefficients for the five indices are in Table ~\ref{tab:calibrations}.

\begin{table}
\caption{Coefficient of the equation $Log(Z/Z_{\odot})= \alpha + \beta EW + \gamma EW^2$
for each index considering the real continuum.}             
\label{tab:calibrations}      
\centering          
\begin{tabular}{cccc}     
\hline\hline       
Index & $\alpha$ & $\beta$ & $\gamma$ \\ 
\hline                    
F1370 &  -2.501 & 1.403 & -0.1700\\
F1425 &  -2.003 & 1.203 & -0.1521\\
F1460 &  -2.023 & 1.251 & -0.1631\\
F1501 &  -2.152 & 2.324 & -0.5329\\
F1978 &  -2.051 & 0.388 & -0.0122\\

\hline                  
\end{tabular}
\end{table}

\begin{table}
\caption{Coefficient of the equation $Log(Z/Z_{\odot})= \alpha + \beta EW + \gamma EW^2$
for each index, considering the pseudo continuum.}             
\label{tab:pscal}      
\centering          
\begin{tabular}{cccc}     
\hline\hline       
Index & $\alpha$ & $\beta$ & $\gamma$ \\ 
\hline                    
F1370 &  -2.897 & 2.346 & -0.4208\\
F1425 &  -2.138 & 2.116 & -0.4424\\
F1460 &  -2.183 & 2.336 & -0.5212\\
F1501 &  -2.543 & 4.800 & -1.968\\
F1978 &  -2.774 & 1.070 & -0.0936\\

\hline                  
\end{tabular}
\end{table}

\subsection{The feature SiII at 1533$\lambda$}
\label{1533}
In addition to the two new indicators found, 
we also studied the SiII $\lambda$1533 feature,
defined as the EWs integrated between 1530 and 1537 \AA~.
This line seems to work very well as indicator:
apart from satisfying the necessary conditions to consider it as good index (as discussed in Sec.~\ref{ind}), 
this line is in fact very deep respect to the others and defined
in narrow wavelength range of 7 \AA~.

Unfortunately, the SiII  is often detected also in emission:
see e.g. the  very high signal to noise ratio spectrum presented by Shapley et al. (2003). 
This emission appears to be very weak, and in low resolution
spectra it may difficult to recognize it.
In addiction, this is  a fine line structure,
(Pettini et al. 2004): although this feature is normally 
photospheric, in  dense environments it can be interstellar as well.
Therefore we suggest to use it with caution, and only in the spectra where this emission is
absent or weak enough to not compromise  the measurements of the EW.

We nevertheless present the calibrations 
obtained for this feature, and discuss the results found for the observed 
galaxies using it as metallicity indicator.

The obtained calibrations are:

\begin{equation}
 \label{equ2}
Log(Z/Z_{\odot})= -1.845 + 1.332  EW - 0.2077 EW^2
\end{equation}
in the case of real continuum, and

\begin{equation}
 \label{equ3}
Log(Z/Z_{\odot})= -1.893 + 1.717  EW - 0.339 EW^2
\end{equation}
 in case of ``pseudo-continuum''.

The results found using this feature are  in Sec.~\ref{res}.

\subsection{Calibration uncertainties}
In previous sections  we defined two new photospheric lines sensitive 
to stellar metallicity and independent to the other stellar parameter (such ad age and
IMF), $\sim1460$\AA~ and  $\sim1501$\AA, and we recalibrate with updated stellar libraries the stellar features at $\sim1370$\AA~ 
$\sim1425$\AA~ and at $\sim1978$\AA~ proposed by Leitherer at al. (2001) and Rix et al. (2004) to derive
the stellar metallicity.

Both the old and the new metallicity indicator suffer of significant uncertainties.

{\it Starburts99} consider two different cases of star formation: an instantaneous 
burst and a continuos star formation at constant rate:
in the former the spectrum changes rapidly with time, in the latter the equilibrium 
is reached after few Myr, and then the spectrum changes little with the time. 
In case of one single burst we expect to observe galaxies  where the O-B stars
are already died, however galaxies with  continuos star formation rate 
are  dominated by bright and hot massive stars which determine the UV continuum (see Adelberger et al. 2004).
Therefore, we decided to consider the continuos star formation rate to create models 
with  {\it Starburts99} because it seems to be the better description for most
star forming galaxies.

Another critical issue in this kind of work is the determination of the continuum.
It is worth noticing that, unlike what was done in other similar works (see Rix et al. 2004) the
values of EW are relative to the real, theoretical continuum. 
However, often low- and medium resolution spectra
do not have many spectral regions free from absorption lines, and the estimate of such a  continuum 
is not straightforward. For this reason it is common to define a ``pseudo'' 
continuum: this is derived  by fitting  a spline curve through the mean flux in some spectral windows 
 relatively free from emission and absorption lines, and a normalized spectrum is obtained dividing by this fit . 
This definition tends to  underestimate  the real continuum because no spectral
window is totally free from absorptions, but
using the same definition of ``pseudo'' continuum for both calibrations  and 
observations, this uncertainty tends to cancel out.\\
When observing faint, high-redshift galaxies,  this method  is difficult to use, 
either because not enough
signal-to-noise ratio is present in the narrow wavelength ranges (1-2\AA, Rix et al. 2004) used
to define the ``pseudo'' continuum, or because bright sky-lines present at $\lambda>7000$\AA~
can mask these regions. As a consequence, each spectrum needs different recipe to compute the 
continuum according to the wavelength range covered and to the signal-to-noise ratio in each point.
Because in same cases is easier use the ``pseudo'' continuum and in other cases
is better to estimate the real   continuum, we provide both the calibrations.

The coefficients for the calibrations using the definition
of the ``pseudo-continuum'' by Rix et al. (2004) (see Table 3 in their paper for the regions used
to define the continuum)  and using the last version
of {\it Starburts99}  are in  Table~\ref{tab:pscal}.
Fig.~\ref{cal1} (right panel) show the relations obtained between metallicity and EW.  

A side effect of these two different procedures is that these calibrations have no or very weak dependency
on spectral resolution, as far it is high enough to well sample the region of interest, i.e.,
10\AA~ rest-frame. Only the F1978 index depends strongly on resolution,
as already noticed by Halliday et al. (2008). In addition this index 
depends also on the IMF, as shown  Fig.~\ref{cal1}.

It is worth noticing that these latter calibrations are more easy to use because the
 definition of the ``pseudo continuum'' is unambiguous: 
in fact only some tiny and defined regions
are used to defined the  ``pseudo continuum'', and not the entire continuum is
considered, as in the other case.
Therefore, we recommend to use the ``pseudo continuum''
 in case of spectra with high signal-to-noise
ratio, where the continuum is less affected by bright sky lines and emission or
absorption interstellar lines,  
and in case of low signal-to-noise spectra we suggest to use the first relations
and  define the real continuum for each spectrum in order to select regions not 
affected by strong, bright sky lines.




\begin{table*}
\caption{Properties of objects observed with FORS: Col. 1 object name in the MUSIC catalog (Grazian et al. 2006), 
Cols. 2,3,
coordinates (J2000), Col. 4 R-band magnitude, 
Col. 5 spectroscopic redshift, Col. 6 the signal to noise ratio, Col. 7 Stellar Mass and Col. 8  gas metallicity. }             
\label{tab:OBJ}      
\centering      
\begin{tabular}{cccccccc}     
\hline\hline       
ID    & RA(J2000)  & DEC(J2000)  & Rmag &  $z$  & SNR\tablefootmark{a}  & $LogM$ & $12+Log(O/H)_{gas}$\\ 
\hline                    
CDFS-12631 & 03 32 18.1 & -27 45 19.0 & 24.72     &  3.709 &   7    &  $9.84^{+0.15}_{-0.07}$ & $8.22^{+0.18}_{-0.14}$\tablefootmark{b}\\  
CDFS-9313  & 03 32 17.2 & -27 47 54.4 & 24.82     &  3.654 &   8    &  $9.34^{+0.17}_{-0.23}$ & $7.95^{+0.25}_{-0.23}$\tablefootmark{b}\\
CDFS-6664  & 03 32 33.3 & -27 50 07.4 & 24.80     &  3.797 &   8    &  $8.98^{+0.20}_{-0.09}$ & $7.63^{+0.38}_{-0.29}$\tablefootmark{c}\\
CDFS-5161  & 03 32 22.6 & -27 51 18.0 & 24.96     &  3.660 &   4    &  $9.73^{+0.22}_{-0.23}$ & $7.69^{+0.23}_{-0.38}$\tablefootmark{b}\\
CDFS-4417  & 03 32 23.3 & -27 51 56.8 & 23.42     &  3.473 &   14   & $10.38^{+0.15}_{-0.04}$ & $8.55^{+0.09}_{-0.10}$\tablefootmark{c}\\
\hline                  
\end{tabular}
\tablefoot{  
\tablefoottext{a}{Signal-to-noise ratio in the wavelength used\\}
\tablefoottext{b}{ Troncoso et al. in preparation } 
\tablefoottext{c}{ Maiolino et al. 2008}.
}
\end{table*}

\section{Stellar metallicity in high redshift galaxies in the AMAZE sample.}

In this section we  apply the method and the relations presented above
to a sample of five
galaxies at $z\sim3.3.$, in order to find, for the first, the stellar
mass-metallicity relation at this redshift.

\subsection{Observations}
\label{sec:obs}
For the present investigation we selected a sub-sample of AMAZE galaxies
at $z\sim3.3$ for which gas metallicities and stellar mass were 
already accurately measured by Maiolino et al. (2008) and 
Trocoso et al. (in preparation, see Table~\ref{tab:OBJ}). 
The masses presented in Table~\ref{tab:OBJ} are slightly	
different from those presented in Maiolino et al. (2008)
because the calculation was improved by using new Spitzer IRAC photometry available.
The observed galaxies have been originally selected in the GOODS-MUSYC sample (Grazian et al. 2006),
and are all in same field allowing for multi-object spectroscopy.\\
Observations of the galaxies were obtained in service
mode in three runs (Nov-Dec.~2008, Nov-Dec.~2009 and Sept-Dec.~2010) 
using FORS2 (FOcal Reduced and low dispersion Spectrograph 2, Appenzeller et al. 1992) 
at the ESO VLT (UT4), under seeing conditions of about 0.8
for each run. The extended
multi-object mode (MXU) was used to optimize the number and placement
of targets in the masks. 

To assure a homogeneous sample the observations were
carried out using a standardized set-up. The slit width was equally 
set to 1 arcsec, corresponding approximately to the size of
typical high-redshift galaxies under average atmospheric conditions
at Paranal. For all observations, 
the low resolution grism 300I was used. This grism covers a 
range of the CCD sensitivity ($6000-11000$ \AA~) with a relatively
high efficiency, reaching its maximum at around 8600 \AA~.
The total exposure times was 37h.\\

\subsection{Data reduction}

The data reduction was performed using the ESO FORS-pipeline.
The default reduction of spectroscopic science data is as follows: 
the data are first corrected for bias, then an extracted mask, 
containing the positions of the long slit and the spatial 
curvatures of the spectra, is
applied to the science data; they are then flat-fielded and re-mapped eliminating
the cosmic-ray
 and the optical distortions. Afterwards they are rebinned to constant wavelength
steps. The wavelength calibration is adjusted using sky emission lines:
this allows to correct for shifts between night-time science and day-time
calibration data.\\
The sky background is obtained using a median of the pixel free from emission lines,
and the sky is subtracted before remapping, i.e. when the spectra are still 
in the original CCD coordinate system.
The sky is determined with a robust linear fitting, which allows for a linear 
spatial gradient in the background.\\
Flux calibration was performed using spectra
of spectrophotometric standard stars obtained each night.
These spectra were reduced with the pipeline, as for the science data,
and used to convert the ADUs into flux units. The 
observed spectra of the standard stars are divided by the corresponding
stellar spectra taken from the literature in order to obtain
response curves and calibration factors. Representative correction curves
are created for each run combining the individual response
curves and smoothing the result by a spline interpolation.
Finally we applied the correction curves and calibration factor
 obtained at all the observed spectra. \\
At the end we obtained 120 flux calibrated spectra for each galaxy.\\
We combined all the reduced data and finally
we used the IRAF \footnote{IRAF (Image Reduction and Analysis Facility) 
is distributed by the
National Optical Astronomy Observatories, which are operated by the
Association of Universities for Research in Astronomy, Inc., under co-
operative agreement with the National Science Foundation.}
 task apall, with the aperture size of 5 pixel, to extract
one-dimensional spectra for the objects.
The list of the sample of galaxies selected for this project are listed
in Table~\ref{tab:OBJ}. The table lists the properties of the
objects: ID, RA, DEC, R magnitude,  spectroscopic redshift, 
signal to noise ratio of the summed spectra, stellar mass and  gas phase metallicity.

Because of the low signal to noise ratio of some reduced data,
we can not use all the spectra for our analysis. After a careful
visual inspection, we removed from our sample the galaxy CDFS-5161
because its spectrum presents too low signal-to-noise ratio 
to detect the single lines.  Moreover, we combined the three
galaxies  CDFS-12361, CDFS-9313, and CDFS-6664 in order to improve the signal
of the single spectra (hereafter we call it CDFS-comb spectrum). The spectrum of the CDFS-4417 galaxy was good 
enough to be used individually.\\

\subsection{Continuum level and EWs}
\label{cont}

Defining the continuum is a critical step in measuring the EWs.
First, the signal-to-noise ratio of our spectra is generally limited (see Tab. \ref{tab:OBJ})
and changes significantly with wavelength because of the presence of sky-lines. 
Second, at our resolution most of the spectrum is  affected by absorption lines. 
Taking into account these two problems, we fit the continuum by using a third- or fifth-order polynomial,
depending on the continuum shape, by using only regions that are expected to be free from deep 
absorption lines. To define the continuum,
 we use the {\it Starburst99} spectra, excluding all the regions 
where absorption larger than 5\% are expected for solar metallicities. 
This method defines a continuum which is slightly underestimated. 
The amount of correction needed can be measured by applying the same technique to theoretical 
spectra, where the continuum level is well know. We obtain that our fitting method  defines 
a continuum level that is
3\%$\pm$2\% below the real one, depending on the spectrum used. This means that the continuum 
bias can be removed by
multiplying the normalized spectra by 1.03.
A series of simulations were also used to estimate the uncertainties on the continuum 
level in each target galaxy.
Random gaussian noise was added to the {\em Starburst99} spectra to obtain the same 
signal-to-noise observed in 
each spectrum, and the continuum was fitted, repeating the procedure many times. 
We obtain that the continuum  level is uncertain of about 
5\% for CDFS-4417, and 10\% for the combined
 spectrum, and that this uncertainty is fairly
 constant with wavelength.

The EW are measured with respect to this fitted continuum, after excluding regions 
affected by bright sky lines. 
The uncertainties on the EWs derive from both the 
poisson noise on the used pixels, and on the uncertainties on the 
continuum level described above.  The latter contribution is dominant 
in all cases, and can become very large for F1978 which is 85\AA~ wide. 
On the contrary, F1501 index is define on the narrowest wavelength range and
is less affected by this contribution. As a result, usually F1501
is the most reliable index for our galaxies.

\begin{figure*}
   \centering
   \includegraphics[width=12cm,angle=270]{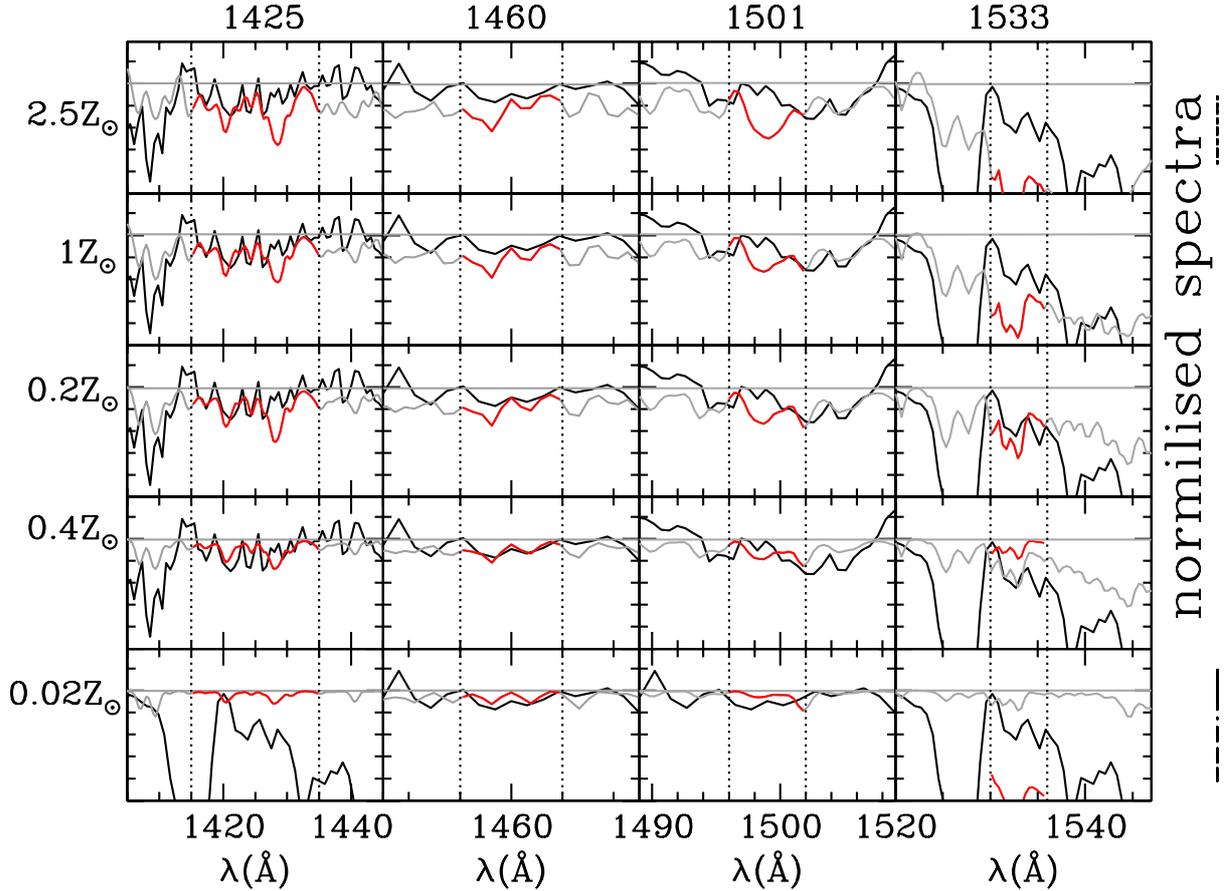}
      \caption{
        Comparison of the observed spectrum of the object  CDFS-4417 at $z= 3.47$
        (black line) with the theoretical spectra (dotted grey line) produced by  {\it Starburts99}
        for five different metallicities, for the each index used to compute 
        the stellar metallicity. In red are highlighted the metallicity indicators.
        All the spectra are normalized by the fitted continuum.}
         \label{1501}
   \end{figure*}
  
\begin{figure*}
   \centering
   \includegraphics[width=12cm,angle=270]{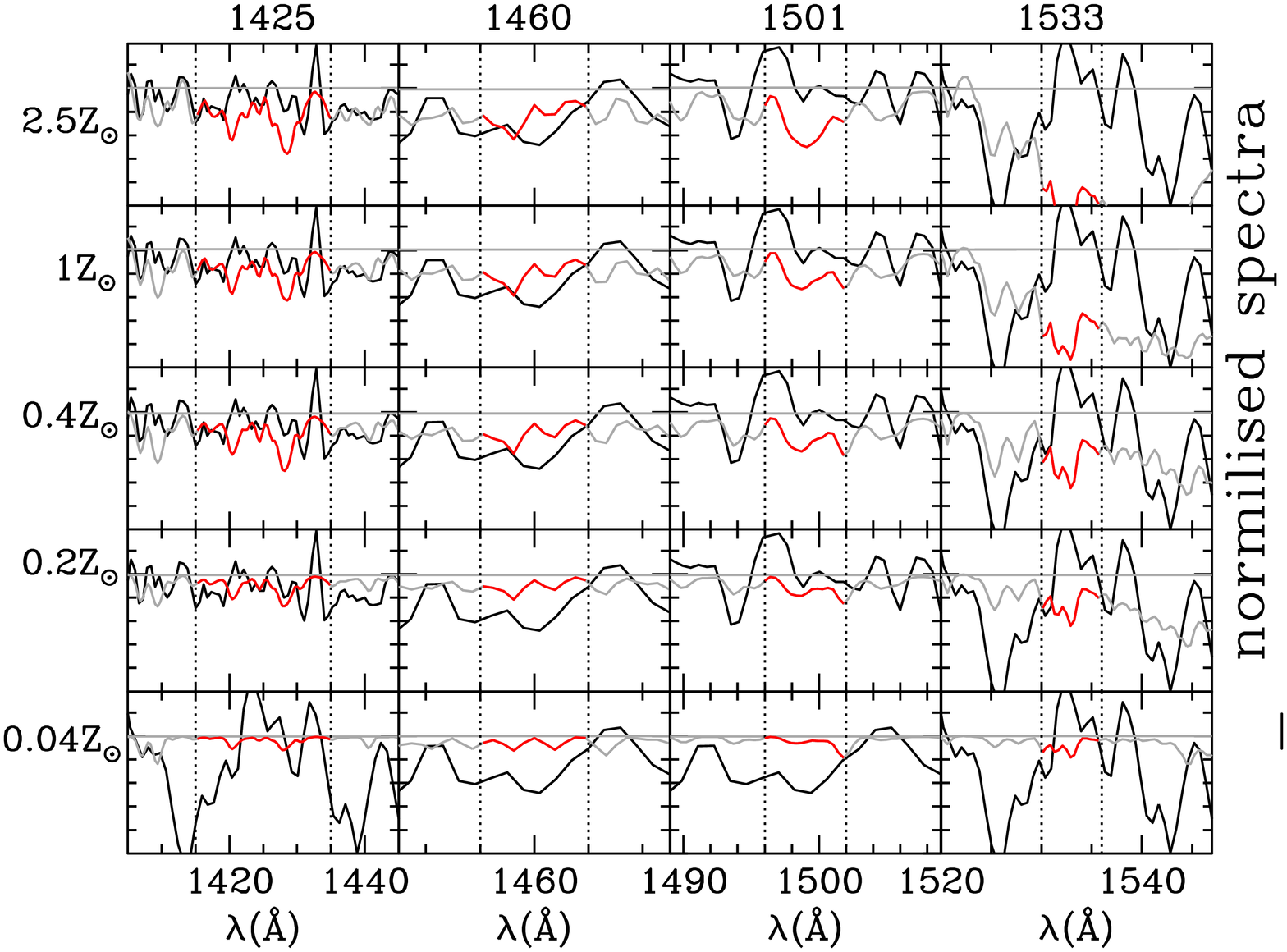}
      \caption{
        Comparison of the observed combined spectrum of the object  CDFS-comb at $z= 3.71$
        (black line) with the theoretical spectra (dotted grey line) produced by  {\it Starburts99}
        for five different metallicities, for the each index used to compute 
        the stellar metallicity. In red are highlighted the metallicity indicators.
        All the spectra are normalized by the fitted continuum.
              }
         \label{789}
   \end{figure*}

  
\subsection{Results}
\label{res}
The EWs measured as described above are compared with the calibrations 
in Tab. \ref{tab:calibrations} to obtain the stellar metallicities.
We will not consider neither F1370,
which is not measurable in our spectra of limited signal-to-noise,
nor F1978, which, for our redshifts, falls in wavelength regions covered by many sky lines.
Therefore for the observed galaxies we can use only F1425, F1460, and F1501.

Fig.~\ref{1501} and Fig.~\ref{789} show the comparison of the observed
spectrum of CDFS-4417 (solid line) and the composite spectrum  CDFS-comb (solid line)
with the theoretical model 
(dotted line), computed for five different metallicities 
for the indices considered to compute the stellar metallicity.
Among the five models plotted in the figures, the metallicities
that most closely match the observations are 0.2$Z_{\odot}$,
in all the spectra regions considered.

This agreement can be made more quantitative using the equations reported in  Table \ref{tab:calibrations}
and computing the metallicity for the observed objects.
In Table  \ref{tab:RES} we present the value found for
each available features with the associated error calculated
applying the theory of propagation of  the error, i.e.
we propagated the uncertainties both in the 
observations and the models to get the final value.
The last column of the Table \ref{tab:RES} represents 
the final value of metallicity from weighted average
of the value of the single features and the associated 
weighted error.

Because in our spectra we did not observe any emission
at 1533$\lambda$, 
we calculated the metallicity of the spectrum 
CDFS-4417 applying the Equation~\ref{equ2},
and we found a value of  $Z=0.26Z_{\odot}\pm0.12$.
This value is in good agreement with what found
using the other indicators (0.23$\pm0.22$), and the associated formal error is lower.
This suggests that it is possible to use the  SiII at 1533$\lambda$
as metallicity indicator when the spectra is free from the 
Si II* fine-structure emission line. Tthe final value of the stellar metallicity does not change
even computing the weighted average using also this feature, where possible.

The gas phase metallicity of the CDFS-4417  was taken from Maiolino et al. (2008),
while for the  combined FORS spectrum 
we combined rest frame optical spectrum in the same way we did for the rest frame UV one, 
in order to derive consistent properties.
 We measured the gas phase metallicity with the $R_{23}$ using the line 
fluxes measured on the combined spectrum.

\begin{table*}
\caption{Metallicity measured in the observed galaxies with the errors computed used all the available indicator,
and weighted average.}             
\label{tab:RES}      
\centering          
\begin{tabular}{c c c c c}     
\hline\hline       
ID         & F1425         & F1460 & F1501 & Weighted\\
           &               &       &       & Average\\  
\hline
CDFS-4417  & 0.14$\pm0.44$ & 0.17$\pm0.47$ & 0.30 $\pm0.32$&    0.23$\pm0.22$       \\
CDFS-comb        & 0.13$\pm0.74$ & 0.37$\pm0.62$ &  --  &     0.24$\pm0.45$       \\
\hline                  
\end{tabular}
\end{table*}

\section{Other data from literature}
\label{litt}

To enlarge the sample we collected from  the literature
other rest frame UV spectra suitable for this kind of work.\\
To compute the gas-phase metallicity we consider one of the most frequently 
used metallicity diagnostics, the $R_{23}$ parameter, defined as:
$$R_{23} = \frac{F({\rm [OII]}\lambda3727) + F({\rm [OIII]}\lambda4959 + F({\rm [OIII]}\lambda5007) } { F({\rm H_{\beta}}\lambda4861) }$$

where F([O II]$\lambda$3727), F([O III]$\lambda$4959) and so on denote the emission-line fluxes of [O II]$\lambda$ 
3727, [O III]$\lambda$4959 and so on, respectively.
 The  $R_{23}$ was proposed by Pagel et al. (1979), and its calibration to the oxygen abundance 
has been improved by both photoionization model calculations (e.g., McGaugh 1991; Kewley \& Dopita 2002),
and empirical calibrations (Nagao et al. 2006). For consistency with the FORS galaxies discussed above,
and to compare with the data of Mannucci et al. (2010), we used the most recent  $R_{23}$ calibration
provided by Maiolino et al. (2008).
\\

\subsection{The Cosmic Horseshoe}

The spectrum of Cosmic Horseshoe, 
a gravitationally lensed galaxy at $z=2.38$,
was analyzed by Quider et al. (2009).
They measured the EW of the F1425 and
 following the definition
and obtained the metallicity $Z=0.5Z_{\odot}$.
Their value of the EW are obtained considering
the pseudo continuum as defined in Rix et al. (2004).
To make a consistent comparison of the Cosmic Horseshoes metallicity
with the other galaxies, we measured
the EWs using the definition of the real continuum described in \ref{cont},
and computing the metallicity with the equation \ref{equ}.
With our procedure the metallicity found is  $Z=0.37Z_{\odot}$, i.e.  $12+\log(O/H)= 8.26\pm029$.\\
For consistency, 
we recomputed the gas phase metallicity for this
galaxies using the $R_{23}$ index with the emission line fluxes taken 
from Hainline et al. (2009) and
the metallicity  calibration of Maiolino et al. (2008).
In this way we obtained $12+\log(O/H)=8.48\pm0.1$.\\
We measured, for the first time, the stellar mass 
of the Cosmic Horseshoe.
We obtained the IRAC photometry at 3.6 $\mu$m and
4.5  $\mu$m  by using Spitzer archive images and   
using a photometric aperture of 8 arcsec and subtracting the flux from the central lensing galaxy. 
The U, G and I photometry were taken from  Belokurov et al. (2007).
We used the Hyperzmass code (Pozzetti et al. 2007 and Bolzanella et al. 2000),
with Bruzual \& Charlot (2003) libraries, assuming a  Chabrier IMF (Chabrier et al. 2003) with an upper mass limit of 100$M_{\odot}$,
smooth exponentially decreasing Star Formation Histories (SFHs) with time scale $\tau=[0.1,\infty]$ and age $t=[0.1,20]$,
deriving  a mass of $Log(M/M_{\odot})=10.56\pm0.19$, corrected for the magnification.

\subsection{The Cosmic Eye}

The lensed galaxy Cosmic Eye at $z=3.075$, has been
studied by Quider et al. (2010).
They did not use the method explained by Rix et al. (2004) to compute the metallicity
because of the heavy contamination of absorption lines
in the region of  the F1425 index that could compromise the definition
of the continuum. They gave just  an indication of the metallicity,  $Z\sim0.40Z_{\odot}$,
comparing the observed spectrum  with the models. 
Computing the EW of the other indices and applying our calibration,
 we found  metallicity  $Z=0.30Z_{\odot}$,
i.e. $12+\log(O/H)=8.16\pm0.28$.\\
The gas phase metallicity is $12+\log(O/H)=8.60\pm0.11$, computed taken the value of $R_{23}$ from Stark et al. (2008)
and applying it in the metallicity calibrations of Maiolino et al. (2008).\\
The stellar mass of the Cosmic Eye derived from SED fitting is  $Log(M/M_{\odot})=9.55\pm0.14$ (Troncoso et al. in preparation).

\subsection{MS 1512-cB58}
The last spectrum that we obtained
belong to the MS 1512-cB58 (Pettini et al. 2000), 
a gravitational lensed galaxies at $z=2.72$ that, thanks to the magnification,
presents a spectrum with very high signal to noise ratio.
As for the previous galaxies, we measured the EWs for all the defined indicators
 on the spectra using the
real continuum and and we obtained a stellar metallicity from 
 equation \ref{equ}  of $Z=0.44Z_{\odot}$, i.e.  $12+\log(O/H)=8.33\pm0.25$.\\
To compute the gas phase metallicity
we took the emission line fluxes in Teplitz  et al. (2000), and applied the 
$R_{23}$ calibration of Maiolino et al. (2008). The result is
 $12+\log(O/H)=8.35\pm0.15$.\\
The stellar mass of MS 1512-cB58 is  $Log(M/M_{\odot})=8.94\pm0.15$ (Siana et al. 2008).

\subsection{Other galaxies}

For five additional high redshift galaxies it was not possible to
obtain the spectra,  but only  the value of the EWs from the literature.
This is the case of the lensed galaxy  8 o'clock arc at $z=2.73$
studied by Dessauges-Zavadsky et al. (2010), as well as for the four galaxies
of the Fors Deep Field with redshift $2.3<z<3.5$ presented in Mehlert et al. (2006).
Both these papers provide the F1425 index, as defined by Rix et al. (2004),
based on  the ``pseudo continuum''.\\
For the lensed galaxy 8 o'clock arc we found that the stellar metallicity is 
$Z=0.52Z_{\odot}$, i.e. $12+\log(O/H)= 8.40$.\\
 Again, to be consistent with the other
results, the gas phase metallicity  was computed with the $R_{23}$ index with the emission line fluxes 
taken from   Finkelstein et al. (2009) and using the calibrations
of Maiolino et al. (2008). We obtained  $12+\log(O/H)= 8.48\pm0.1$.\\
The stellar mass of this galaxy is  $Log(M/M_{\odot})=10.25^{+2.22}_{-0.68}$ (Richard et al. 2011).
We can not give the errors  for the stellar metallicity of this galaxy
because Dessauges-Zavadsky et al. (2010) 
did not give any indication of the uncertainties relative to the EWs measurements.\\
In the same way, we calculated the stellar metallicity for four FORS Deep Field galaxies 
studied by  Mehlert et al. (2006). It was not possible to compute the gas phase metallicity
because the emission line fluxes for these galaxies are not available.
We reported the results in Fig.~\ref{MZ}, where we  highlighted	
these objects with black crosses. The masses of these galaxies were provided by Drory et al. (2005 and private communication).

\begin{table*}
\caption{Comparison between the stellar metallicity
obtained with this work and gas phase metallicity.
In all the cases we assume a solar value of $log(Z/Z_{\odot}= 12 + log(O/H) - 8.69$ (Allande Prieto et al. 2001).
The Mass is $\log(M/M_{\odot})$.}             
\label{tab:comp}      
\centering          
\begin{tabular}{c c c c c c}     
\hline\hline       
ID         & Mass & redshift & stellar metallicity   & gas metallicity\\  
\hline                    
CDFS-4417    &  10.38 & 3.47 & 8.05$\pm0.22$ & 8.46$\pm0.11$\tablefoottext{a}\\
CDFS-comb    &   9.38 & 3.71 & 7.92$\pm0.24$ & 7.98$\pm0.15$\tablefoottext{b}\\
horseshoe    &  10.56 & 2.37 & 8.26$\pm0.29$ & 8.48$\pm0.11$\tablefoottext{c}\\
8oclock      &  10.25 & 2.73 & 8.40$   ---   $ & 8.42$\pm0.10$\tablefoottext{d}\\
Cosmic eye   &   9.60 & 3.07 & 8.36$\pm0.20$ & 8.60$---$\tablefoottext{b}\\ 
MS 1525 cB58 &   8.94 & 2.72 & 8.33$\pm0.25$ & 8.35$\pm0.15$\tablefoottext{e}\\
FDF-3173     &   9.96 & 3.27 & 8.54$---$     & $-------$& \\
FDF-3810     &  10.95 & 2.37 & 8.90$---$     & $-------$ & \\
FDF-5903     &  11.01 & 2.77 & 8.24$---$     & $-------$ & \\
FDF-6934     &  11.31 & 2.44 & 8.57$---$     & $-------$ & \\ 
\hline                  
\end{tabular}
\tablefoot{ The gas metallicity was calculated using the calibration
of Maiolino et al. (2008), and the emission line fluxes taken from
\tablefoottext{a}{Maiolino et al. (2008) }
\tablefoottext{b}{Troncoso et al. in prep.} 
\tablefoottext{c}{Hainline et al. (2009)} 
\tablefoottext{d}{Finkelstein et al. (2009)}
\tablefoottext{e}{Teplitzt al. (2000)}.
respectively.}
\end{table*}

\section{Comparison between stellar and gas-phase metallicities}

In this section we compare the stellar metallicity obtained using the empirical calibrations
described above with the gas phase ones.

Small differences are expected between the stellar metallicities as measured using
UV absorption features and the gas phase metallicities obtained by strong optical
emission lines. Nevertheless, if the galaxy is experiencing a rapid metallicity evolution, 
differences could be related to the longer
lifetimes of the star dominating UV emission with respect to the more massive stars dominating line emission.
In fact, the stars responsible of the UV emission are young, hot O-B stars (with a
life time of $\sim10^{6}-10^{7}$ yr), 
which  were formed from interstellar gas with very
similar properties of the one seen in emission given their short lifetime.

On the other hand, larger differences are expected using optical absorption features,
dominated by longer lived stars (e.g. Lick indices, 
as found in Galazzi et al. (2005) and Panter et al. (2008) for local SDSS galaxies).

In Table \ref{tab:comp} we report the comparison between stellar and  gas
phase metallicity for the objects that we discussed in the previous sections.
The differences between the stellar and
the gas-phase metallicity are plotted in Fig.~\ref{dif}.
For each galaxies the error are computed by combining the errors
from the gas-phase and stellar metallicities in quadrature.

As expected, we do not find large discrepancy between the two quantities:
 the average difference is -0.16 and the uncertainty is 0.14,
therefore the difference has a significance of about 1.1sigma. 
This does not allow us to claim any significant difference,
given the large systematic uncertainties associated with both methods.
Also Halliday et al. (2008), analyzing star forming galaxies at $z\sim2$  found 
the stellar metallicity lower by $\sim0.25$dex than the gas phase one measured for galaxy with similar stellar mass and redshift.
It is worth noticing that, since  Halliday et al. (2008) are not comparing the  metallicities of the same galaxies,
some selection effects present in one or both data sets could affect the results.
To verify if any difference is real larger sample of galaxies are clearly needed.

\section{The stellar mass-metallicity relation}

Our data allow for the first time the study of the stellar mass-metallicity relation
at high redshift.

\begin{figure}
   \centering
   \includegraphics[width=9cm]{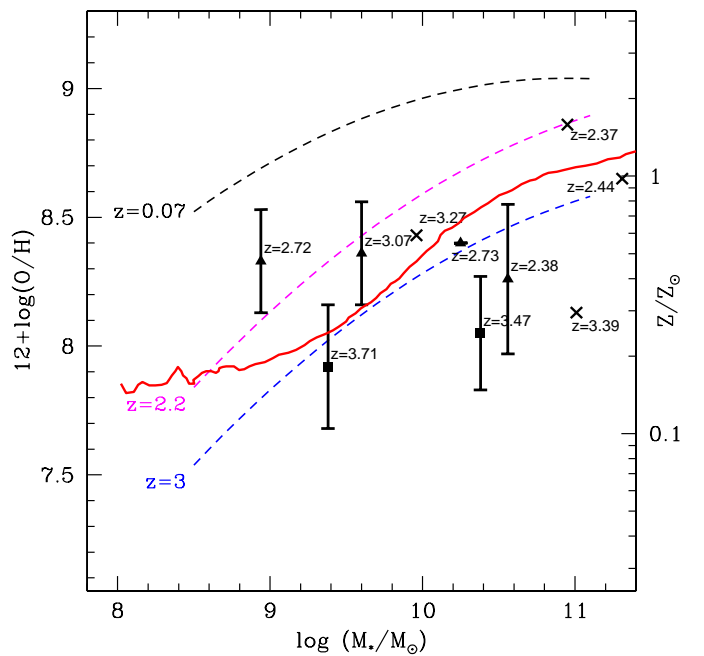}
      \caption{
        Stellar mass-metallicity relation for the FORS galaxies at $z\sim3.3$ discussed in this work (black squares).
        The black triangles 
        represent the stellar metallicity for the lensed galaxies from the literature (see text), and 
        the black crosses are the stellar metallicity for the galaxies at $z\sim3$ taken by Mehelert et al. (2006).
        For comparison, we draw the mass-metallicity relations for the gas component:
        the black line is the MZR obtained by Tremonti et al. (2004) at $z=0.07$,
        the magenta line was found by Erb et al. (2006) at  $z=2,2$, and the blue dotted line
        shows the relation from the LSD and AMAZE inferred by 
        Mannucci et al. (2009) at $z\sim3.5$. The red line is stellar mass-metallicity relation
        found by Panter et al. (2008) in local universe. 
                  }
         \label{MZ}
   \end{figure}

Fig.~\ref{MZ} shows the position and the relative error of the 
observed galaxies (black squares)
at $z>3$ presented above in the MZ-plane. 
The black triangles represent the stellar metallicity for the lensed galaxies.
The black crosses are the stellar metallicity calculated
for galaxies at $z\sim3$ taken by Mehelert et al. (2006). 
All the quantities are computed in the same way and are consistent.

To compare our finding with the previous studies, in the Fig.~\ref{MZ}
we draw the various mass-metallicity relations found a different 
redshifts  both for stellar and gas phase metallicity. 
For the latter, the black dotted line is the relation obtained by Tremonti et al. (2004) 
in the local universe as derived from the the SDSS, at $z\sim 0.07$,
the magenta dotted line represents the relation obtained by Erb et al. (2006) at $z\sim2.2$,
and the blue dotted line shows the behavior of such relation inferred from the initial
sample of LSD and AMAZE sources at  $z\sim3$ by Mannucci et al (2009).\\
The green dotted line in Fig. ~\ref{MZ} represents the stellar metallicity
relation found by Panter et al. (2008) for local SDSS galaxies.\\

As discussed in the previous section, the derived  stellar metallicities are consistent, 
although slightly  lower, than the gas phase ones. 
The stellar metallicities therefore provide an independent test of the 
reliability of the metal abundances in high redshift galaxies, 
more often obtained for the gas phase only using strong optical emission line ratios.  
In fact, the strong line diagnostics used are indirect tracers, calibrated in local galaxies, 
and depend not only on metallicity, but also on other parameters as ionization conditions and densities, 
which might in principle  be very different in high-z star forming galaxies than in local spirals 
(see e.g. Nagao et al. 2006, Brinchmann et al. 2008). 
Our result therefore confirms the low chemical abundances derived from optical  line ratios at high redshift, 
and the strong evolution in gas phase and young stars metallicity  with respect to local and lower redshift galaxies, 
as already claimed by Maiolino et al. (2008) and Mannucci et al. (2009).
Although the large uncertainties in both stellar and gas-phase metallicity measurements, 
it additionally support the finding of Mannucci et al (2010) and Troncoso et al. (in prep.) that the Fundamental
 Metallicity Relation between mass, metallicity and SFR is evolving at $z\sim3$,
 while no evolution is found at lower redshifts (see also Cresci et al. 2011).

\begin{figure}
   \centering
   \includegraphics[width=9cm]{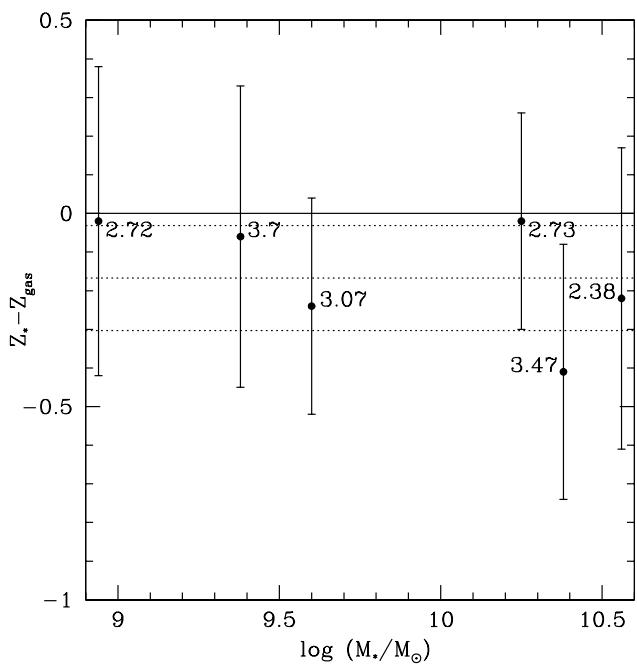}
      \caption{
        Difference between the stellar and the gas-phase metallicity of the galaxies.
        The redshift of each source is reported in the labels. The dotted lines
        are the mean value of the differences and the 1-$\sigma$ deviation.
                  }
         \label{dif}
   \end{figure}

As shown in  Fig. ~\ref{dif}, the stellar metallicities derived for $z\sim3$ galaxies 
are comparable with the ones obtained by 
Gallazzi et al. (2005) and Panter et al. (2008) for local SDSS galaxies.
However the stellar populations that are dominating the spectral features considered in the
two cases to compute the metallicity are different: in fact, in high redshift galaxies 
the rest frame UV spectrum that we observe is dominated by
hot, young stellar populations, while in the local universe
the optical indices used by Panter et al. (2008) are dominated by cold, older stars: 
as noted by Panter et al. (2008), 
their stellar metallicities for the galaxies with a younger stellar populations ($\leq$1 Gyr) 
are better in agreement with the gas phase ones, while larger deviations are present for 
the galaxies dominated by an older population. Therefore, a direct, straightforward  comparison of the two is not possible.
However, the fact that the stellar metallicities of the SDSS sample 
are comparable with both gas phase and stellar abundances  at $z>2.5$ is an indication
that the bulk of the stellar populations in the galaxies investigated by Panter et al. (2008) were formed during this epoch.

\section{Conclusions}

In this paper, we have investigated for the first time the stellar 
mass-metallicity relation at high redshift, $z\sim3$.\\
Using the theoretical spectra
created with the population synthesis code  {\it Starburts99},
we looked for photospheric absorption lines to be used as indicators
of stellar metallicity.\\
First we tested the line indices proposed by
Leitherer et al. (2001) and Rix et al. (2004),
the  F1370, F1425 and F1978 using the last version of {\it Starburts99}, although 
the F1978 shows a strong dependence from the resolution and the IMF.\\
Then we defined two new photospheric lines, 
F1460 and F1501, and we found that these lines are sensitive to the metallicity
and almost independent of the age and the IMF, and therefore useful stellar metallicity
indicators. The F1501 index seems to be the most promising 
because it is defined on the narrowest wavelength range and
less affected by the uncertainties on the continuum definition.\\
We provided the metallicity calibrations, see Fig.~\ref{cal1}, with two different
definitions of the continuum: the first relations are referred to the real continuum,
that we suggest to use in case of spectra with low signal-to-noise ratio, the others
were  obtained using the definition of the ``pseudo-continuum'' provided
by Rix et al. (2004), that we recommend in case of high signal-to-noise spectra,
see Sec.~\ref{mc}.\\
We applied the relations on one galaxy and a composite spectra comprised
of three additional galaxies of the AMAZE sample at
$z\sim3.3$, for which the gas phase metallicity and the galaxy masses
were already know.\\ 
We took from  the literature the 
spectra of eight additional galaxies, and we recompute their stellar metallicity using the new calibrations.\\
At the end we compared the results found with
the gas phase metallicity for each object, see Fig.~\ref{dif}.
The main conclusion  of this work is that within the  errors,
the stellar and the gas phase metallicity are consistent,
although there seems to be a tendency to find stellar metallicity lower 
than the gas phase one by  $\sim0.1dex$,
 as already found by Halliday et al. (2008). 
This result supports the low metal content derived for the gas phase of high-z galaxies from optical strong line ratios, 
as well as an evolution of the Fundamental Metallicity Relation at $z>3$ as found by Mannucci et al. (2010). \\
For the first time, we obtained the stellar mass-metallicity
at redshift $z>2.5$,  see Fig.~\ref{MZ}. We notice that the stellar metallicities found at high
redshift is comparable with  those found by
Panter et al. (2008)  for local galaxies, although the two are not straightforwardly comparable
as in high
redshift galaxies the stellar metallicity are computed for
hot, young stars, while in the local galaxies for cold, older stellar population.\\

In summary, the rest-frame UV is rich in metallicity dependent features, 
which are able to provide a measure of stellar metallicity in high redshift galaxies. 
This represent an independent measure of the chemical abundances in galaxies with respect to the more widespread gas phase metallicities, 
which can provide important constraints to the star formation histories of galaxies in the early Universe. 
Although this technique is currently limited to very bright or lensed galaxies by the high S/N required,
 the advent of next generation of telescopes will
give us much higher quality spectra for high redshift galaxies, and the stellar metallicity
indicators will play a more important role in chemical abundances studies at high
redshift.

\begin{acknowledgements}
      GC acknowledges financial support from ASI-INAF grant  I/009/10/0.
Thanking Chuck Steidel for his spectrum of MS1512-cB58,
and Alice Shapley for her spectra of some lensed galaxies.
We thank Max Pettini and Claus Litherer for useful comment and suggestions.
We are grateful to Drory for providing the masses of the FDF galaxies.

\end{acknowledgements}

\end{document}